\documentclass[twocolumn]{autart}   
\usepackage{color}
\usepackage{cite}
\usepackage{amsmath,amssymb,amsfonts}
\usepackage{graphicx}
\usepackage{textcomp}

\usepackage{graphicx}          
\usepackage{epstopdf}

\begin{document}


\begin{frontmatter}

\title{Optimal filtering for a giant cavity in waveguide QED systems} 


\author[GG1,GG2]{Guangpu Wu},
\author[GG1,GG2]{ Shibei Xue},\ead{shbxue@sjtu.edu.cn}    
\author[GG1,GG2]{Yuting Zhu},          
\author[GG4]{ Guofeng Zhang},
\author[GG5]{ Ian R. Petersen}

\address[GG1]{School of Automation and Intelligent Sensing, Shanghai Jiao Tong University, Shanghai 200240, People's Republic of China}
\address[GG2]{Key Laboratory of System Control and Information Processing, Ministry of Education of China, Shanghai 200240, P. R. China}
\address[GG4]{Department of Applied Mathematics, The Hong Kong Polytechnic University, Hung Hom, Kowloon, Hong Kong Special Administrative Region, China}
\address[GG5]{School of Engineering, Australian National
University, ACT 2601, Australia}

\thanks{This work was supported by the National Natural Science Foundation of China (NSFC) under Grants No. 62273226 and No. 61873162. This work was also financially supported by Quantum Science and Technology-National Science and Technology Major Project 2023ZD0300600, Guangdong Provincial Quantum Science Strategic Initiative No. GDZX2303007, Hong Kong Research Grant Council (RGC) under Grant No. 15213924, and  the CAS AMSS-PolyU Joint Laboratory of Applied Mathematics. }

\begin{abstract}
In waveguide quantum electrodynamics (QED) systems, a giant cavity can be engineered to interact with quantum fields by multiple distant coupling points so that its non-Markovian dynamics are quite different from traditional quantum optical cavity systems. Towards feedback control this system, this paper designs an optimal filter for the giant cavity systems to estimate its state evolution under continuous quantum measurements.
Firstly, the   Langevin equation in the Heisenberg picture are derived, which is a linear continuous-time system with both states and inputs delays resulting from the unconventional distant couplings.
Compared to existing modeling approaches, this formulation effectively preserves the nonlocal coupling and multiple delay dynamic characteristics inherent in the original system.
 In particular, the presence of coupling and propagation delays leads to noncommutativity among the system operators at different times, which prevents the direct application of existing quantum filtering methods.
To address this issue, an optimal filter is designed, in which the delayed-state covariance matrices are computed.
By iteratively evaluating the delayed-state covariance over successive time intervals, the resulting optimal filter can be implemented in an interval-wise backward recursion algorithm. 
Finally, numerical simulations are conducted to evaluate the tracking performance of the proposed optimal filter for the giant cavity. By comparing between the evolutions of Wigner functions of coherent and cat states and the filter, the effectiveness of the optimal filter is validated.
\end{abstract}


\end{frontmatter}


\section{Introduction}\label{sec1}

Quantum filtering and control constitute fundamental problems to enhance the performance of a quantum system in quantum computing \cite{Livingston2022} and quantum metrology~\cite{bg1, Dong2010, Giovannetti2006}.
From a control-theoretic perspective, the central objective of quantum filtering is to optimally infer the state or dynamical characteristics of a quantum system from limited measurement information~\cite{Wiseman2010, Helstrom1976}.
Quantum filtering theory was first proposed by Belavkin in the  1980s~\cite{Belavkin1989}.
Its fundamental idea is to construct a non-commutative quantum extension of classical stochastic processes and filtering theory, thereby enabling classical optimal filtering methods to be applied to quantum filtering problems.
Compared to classical filtering theory, the modeling and analysis of quantum filters must explicitly account for the non-commutativity of operators and the back-action of measurement on system dynamics, which constitutes an essential feature of quantum filtering problems~\cite{Bouten2007, Wiseman1993}.
Based on this, many methods for quantum filtering  have been proposed for Markovian systems, where the variations of the system states depend only on its current state.
These methods include 
Bayesian mean estimation~\cite{Paris2004}
and extended Kalman estimation~\cite{Yu2018}.
These approaches are grounded in quantum probability theory and have led to the systematic development of quantum filtering theory.

In contrast to Markovian quantum systems, the evolution of a non-Markovian
quantum system interacting with an environment exhibiting memory effects depends on not only the current states but also its past states,
thereby preventing the direct application of established quantum filtering techniques.
To filter this kind of systems, a widely adopted approach  is based on augmented modeling, where ancillary systems are introduced to represent the internal modes of a non-Markovian environment so as to represent the original non-Markovian quantum system in an augmented Hilbert space~\cite{Xue2015,Xue2019}. Concretely, the ancillary system can be determined by a spectral decomposition theorem such that the ancillary together with principal systems can be described by a time-invariant equation. Although this approach can deal with the non-Markovian environment with a rational spectrum, it would result in a high-dimensional augmented systems, thereby incurring a substantial computational burden. 
Besides non-Markovian dynamics induced by the environment, the evolution of a controlled quantum system would be non-Markovian resulting from time-delays in a feedback loop~\cite{Kashima2009,Wang2015}. However, in this case, the filter can be a standard Markovian quantum filter since the design of the filter is irrelevant to the delay in the feedback loop according to the separation principle in control theory.
Recently, a new class of non-Markovian dynamics can be constructed in a waveguide quantum electrodynamics dynamical (waveguide QED) systems, where a quantum system can interact with the field in a waveguide by multiple times via distant capacitors or inter-digital transducers (IDT)~\cite{Schuetz2015, Satzinger2018}.
Since bounded states or non-exponential decays has been observed in this kind of systems which would be useful in quantum information processing~\cite{Guo2020, Andersson2019}, it is important to filter out the state of the system for further applications. However, since the variation of the system state depends on its time-delayed state, standard quantum filter theory would not be applicable. In addition, when using the augmented system approach, time-delayed terms would result in a high-dimensional augmented system which would be computational costing. Hence, the problem to design an effective quantum filter for such system would be open and valuable.

 Motivated by this gap, we develop an optimal filtering framework for non-Markovian cavity systems in waveguide QED. We construct a dynamical model in which the Langevin equations are represented as a linear continuous-time system incorporating mixed delays. This modeling approach retains the essential nonlocal coupling and multi-delay characteristics of the original system. Notably, coupling and propagation delays induce noncommutativity among system operators at different times, rendering existing quantum filtering techniques inapplicable. To overcome this difficulty, we propose an optimal filter design framework based on an interval-wise recursive computation of delayed-state statistics, leading to an interval-wise backward recursion algorithm for filter implementation. Finally, numerical simulations are presented to assess the tracking performance of the proposed filter in the original system, where the results confirm the effectiveness of the proposed approach.


In Section \ref{section2}, we derive the Langevin equation for a cavity
coupled to a waveguide at two distant points as well
as its input-output equation.
In Section \ref{section3}, we design our
optimal filter for estimation of the  cavity mode state.
In Section  \ref{section4},  an example is
given to show the effectiveness of our filter.
Finally, conclusions are drawn in Section  \ref{section5}.

\textbf{Notation} For a matrix $A=[A_{s,t}]$, the symbols $A^T$ and $A^\dag$ represent the transpose and Hermitian conjugate of $A$. Given a complex number $a$, $  a^\dag$ represents its conjugate. The matrix $\mathcal{M}(a)$ is defined as
$\mathcal{M}(a)_{s,t}=\left[\begin{array}{cc}a+a^*&i(a-a^*)  \\-i(a-a^*) &a+a^*\\\end{array}\right]$. Given two operators $M$ and $N$, $[M,N]=MN-NM$ is their commutator.
For an operator $p$,  $\langle p \rangle$ represents the expectation values of this operator.
Symbol ${\rm H.c.}$ represents Hermitian conjugate of the corresponding term.
\section{System description  for a giant cavity  system}\label{section2}

\subsection{Hamiltonian of the giant cavity system}

\begin{figure}[hbt]
\begin{center}
\includegraphics[width=8cm]{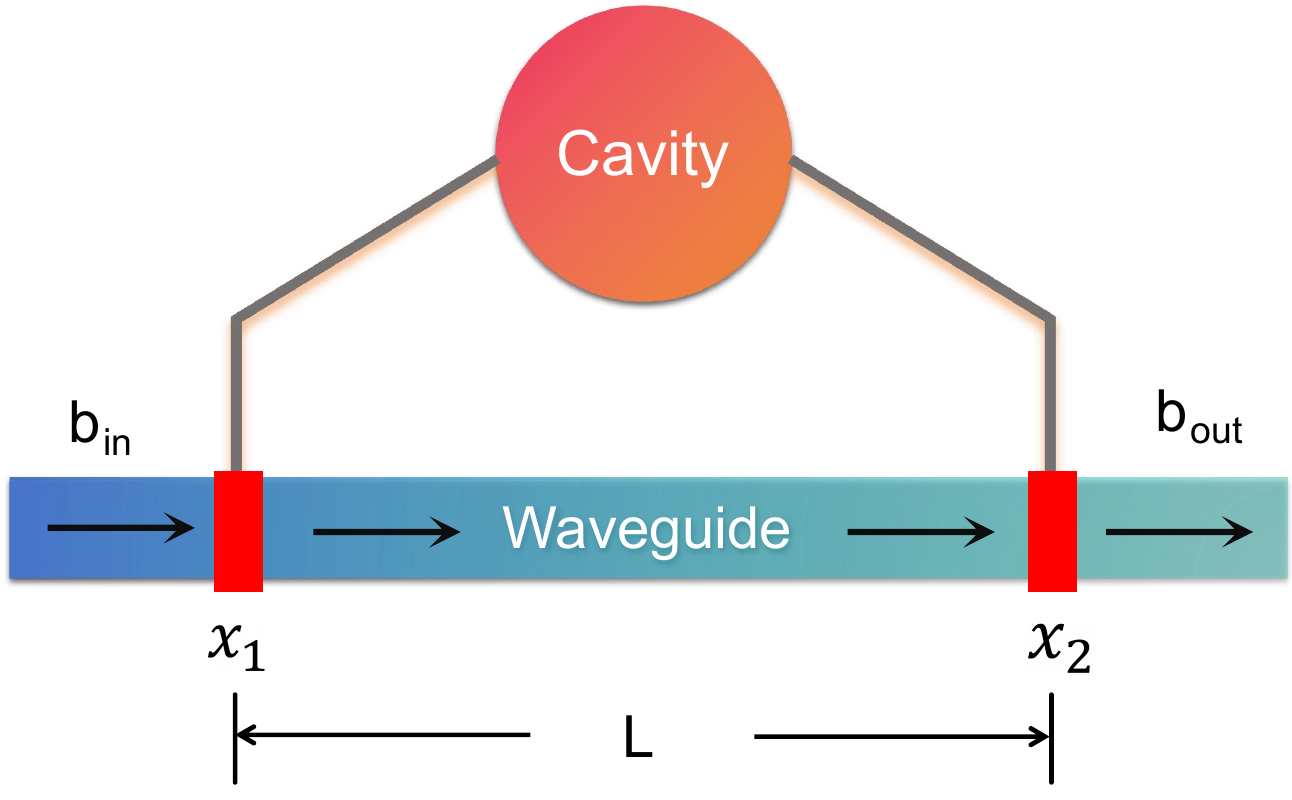}
\end{center}
\caption{Schematic of a cavity coupling to a waveguide at two distant points.}
\label{model1}
\end{figure}

 In waveguide QED, a giant cavity refers to a cavity coupled to a field at distant coupling points which result in the non-Markovian dynamics of the cavity, \cite{Zhu2022b, Kockum2018}. In this paper, we consider a giant cavity interacting with the field in the waveguide at two distant coupling points, which is depicted in Fig. \ref{model1}. An input field $b_{\rm in}$ propagates unidirectionally along the waveguide and interacts sequentially with the cavity at positions $x_1$ and  $x_2$. When the field passes the position $x_2$ after which the field has no interaction with the cavity, we consider it as an output field  $b_{\rm {out}}$.
Unlike the classical single-point-coupled system, due to the finite propagation distance between the two coupling points, the cavity interacts with its own emitted field after a non-negligible time delay, giving rise to intrinsic non-Markovian dynamics \cite{ Du2023}. Physically, such couplings can be realized by multiple capacitors or inter-digital transducers\cite{Zhu2022}. 

To describe the dynamics of the giant cavity system, we first introduce the Hamiltonian formulation of the system--environment model.
The total Hilbert space is given by the tensor product
$
\mathcal{H} = \mathcal{H}_S \otimes \mathcal{H}_E$,
where $\mathcal{H}_S$ and $\mathcal{H}_E$ denote the Hilbert spaces of the cavity mode and the waveguide field, respectively.
Accordingly, the total Hamiltonian can be decomposed as
\begin{equation}
H = H_S + H_E + H_I. \label{1}
\end{equation}
The internal Hamiltonian of the cavity mode is given by
\begin{equation}
H_S = \omega_c a^\dag a, \label{2}
\end{equation}
where $\omega_c$ is the working angular frequency of the cavity.
Here, the annihilation and creation operators $a$ and $a^\dag$ act on the system Hilbert space $\mathcal{H}_S$ and satisfy the canonical commutation relation $[a, a^\dag] = 1$.
The waveguide Hamiltonian takes the standard form
\begin{align}
H_E &= \int_{-\infty}^{+\infty} \mathrm{d}\omega~ \omega\, b_r^\dagger(\omega)b_r(\omega), \label{3}
\end{align}
where $b_r(\omega)$ and $b_r^\dagger(\omega)$ are the annihilation and creation operators of the waveguide field modes with frequency $\omega$.
These operators act on the environment Hilbert space $\mathcal{H}_E$ and satisfy the standard bosonic commutation relations
$[b_r(\omega), b_r^\dagger(\omega')] = \delta(\omega-\omega')$.

The interaction Hamiltonian describing the coupling between the cavity mode and the waveguide field at two spatially separated points is given by
\begin{align} \label{4}
H_I &= \sum_{n=1}^2 \frac{V_q}{\sqrt{v_g}} 
\int_{-\infty}^\infty \mathrm{d}\omega 
\big( b_r(\omega) e^{{\rm i}\omega \tau_n} a^\dag + \text{H.c.} \big),
\end{align}
where $\tau_n$ denotes the propagation-induced time delay associated with the $n$-th coupling point.
We set the first coupling point at position $x_1=0$ and the second one at $x_2=L$, such that $\tau_1=0$ and $\tau_2=L/v_g$, where $v_g$ is the group velocity of the waveguide field.
The parameter $V_q$ denotes the cavity--waveguide coupling strength, which is assumed to be identical at the two coupling points.

\subsection{Motion equation}

With the Hamiltonian, we derive the system's motion equation.
Here, we introduce two important definitions~\cite{Song2016, Guo2017}.


\textbf{Definition 1.}
The input field in the waveguide is written as
\begin{equation} \label{bin}
    b_{\rm in}(t)=\frac{1}{\sqrt{2\pi}}\int^{+\infty}_{-\infty}{\rm d}\omega \, b_r(\omega,0)e^{-{\rm i}\omega t},
\end{equation}
where $  b_{\rm in}(t)$ satisfies $[  b_{\rm in}(t),  (b_{\rm in}(t'))^\dag]=\delta(t-t')$ with $\delta(t-t')$ being Dirac delta function. We have let $b_r(\omega,0)=b_r(\omega)$, where the second variable indicates the initial time $t_0=0$.

This definition introduces the input field as a continuous superposition of right-propagating frequency modes evaluated at the entrance of the waveguide.
It follows naturally from the Fourier representation of traveling-wave bosonic fields and provides a convenient time-domain description for formulating input--output relations which will be given in the next subsection.
The field \( b_{\rm in}(t) \) 
is typically assumed to be in a vacuum or Gaussian state, which ensures the field is quantum white noise.

 We then have the following result.

\textbf{Theorem 1.} Given total Hamiltonian  \eqref{1}, the dynamical equation for the annihilation operator of the giant cavity system is written as
\begin{align} \label{at20}
\dot{a}(t) &= -{\rm i} \omega_c a(t)- \frac{\gamma}{2} \big[a(t) + a(t-T)\big]\nonumber\\
& - {\rm i} \sqrt{\frac{\gamma}{2}} \big[b_{\rm in}(t) + b_{\rm in}(t-T)\big],
\end{align}
 with  $\gamma={4\pi V_q^2}/{ v_g}$ and $T=\tau_2={(x_2-x_1)}/{v_g}={L}/{v_g}$.

\textbf{Proof.} 
In the Heisenberg picture, the cavity mode and field mode evolve as
\begin{equation}
\dot{a}(t) = -{\rm i} \omega_c a(t)- {\rm i} \sum_{n=1}^2 \frac{V_q}{\sqrt{v_g}} \int_{-\infty}^\infty \mathrm{d}\omega \, b_r(\omega) e^{{\rm i}\omega \tau_n}, \label{7}
\end{equation}
\begin{equation}
\dot{b}_r(\omega, t) = -{\rm i}\omega b_r(\omega, t) - {\rm i} \sum_{n=1}^2 \frac{V_q}{\sqrt{v_g}} e^{-{\rm i}\omega \tau_n} a(t), \label{art}
\end{equation}
respectively.
Due to linearity, Eq.~(\ref{art}) can be solved as
\begin{equation}
b_r(\omega, t) = b_r(\omega, 0) e^{-{\rm i}\omega t } - {\rm i} \sum_{n=1}^2 \frac{V_q}{\sqrt{v_g}} \int_{0}^t \mathrm{d}t' a(t') e^{{\rm i}\omega ( t'-t -\tau_n)}. \label{ar}
\end{equation}
Taking Eq. (\ref{ar})  into Eq. (\ref{7}), we have
\begin{align}
\dot{a}(t)& = -{\rm i} \omega_c a(t) - {\rm i} \sum_{n=1}^2 \frac{V_q}{\sqrt{v_g}} \int_{-\infty}^\infty {\rm d}\omega {\Big [}b_r(\omega, 0) e^{-{\rm i} \omega t } \nonumber\\
&- {\rm i} \sum_{{n'}=1}^2 \frac{V_q}{\sqrt{v_g}} \int_{0}^t {\rm d}t' a(t') e^{-{\rm i} \omega (t - t')}e^{-{\rm i}\omega \tau_{n'}}{\Big ]} e^{{\rm i} \omega \tau_n}\nonumber \\
&= -\mathrm{i}\omega_c a(t)-\mathrm{i}\sum_{n=1}^2
\sqrt{\frac{\gamma}{2}}\,
b_{\mathrm{in}}\!\left(t-\tau_n\right)
 \nonumber\\
&\quad
-\sum_{n=1}^2 \sum_{n'=1}^2
\frac{\gamma}{2}\,
a\!\left(t-\tau_{n'}+\tau_n\right)
\theta\left(\tau_n-\tau_{n'}\right)
\label{zhongjian}
\end{align}
where  $\theta(\cdot)$ is the Heaviside step function defined as
$\theta(\tau_n-\tau_{n'}) = 
\begin{cases}
1, & \tau_n-\tau_{n'} > 0, \\[4pt]
\frac{1}{2}, & \tau_n-\tau_{n'} = 0, \\[4pt]
0, & \tau_n-\tau_{n'} < 0.
\end{cases}$
Using  definition \eqref{bin} and  the equality
$ \int^{+\infty}_{-\infty}{\rm d}\omega e^{{\rm i}\omega(t-t')}=2\pi \delta(t-t')$,
we can simplify Eq. (\ref{zhongjian}) as Eq.
\eqref{at20} which captures the dynamic evolution of the cavity mode coupled to the waveguide fields. $\blacksquare$


\textbf{Remark 1.}
The Heaviside step function $\theta(\cdot)$ in Eq. \eqref{zhongjian} is essential because the field propagates uni-directionally such that the state of the system or the field in the downstream position can be influenced by the state of the system or the field in the upstream position, but not vice versa.
Compared with the conventional quantum Langevin equation derived under the Markov approximation, the cavity mode \eqref{7} couples to the waveguide field at multiple spatially separated points.
As a consequence, the interaction explicitly retains phase factors associated with finite propagation times.
In addition, 
due to the presence of distinct coupling delays \( \tau_n \) in Eq. \eqref{art}, the waveguide modes mediate temporally nonlocal feedback from the cavity to itself, giving rise to intrinsic non-Markovian dynamics.

\textbf{Remark 2.}
When the propagation delay vanishes, namely $T=0$, the delayed terms in \eqref{at20} satisfy $a(t-T)=a(t)$ and $b_{\rm in}(t-T)=b_{\rm in}(t)$, and thus \eqref{at20} reduces to the Markovian form $\dot a(t)=-{\rm i}\omega_c a(t)-\gamma a(t)-{\rm i}\,\sqrt{2\gamma}\,b_{\rm in}(t)$ which can be rewritten as the two coupling points  with no-delay equation for a damped cavity mode \cite{Gardiner1985}. Therefore, Theorem 1 can be viewed as a delayed extension of the conventional model, and in the zero-delay limit it recovers the Markovian form.


\subsection{Input-output relation}
 To design a quantum filter, it is required to derive the input-output relation for the system.
 
\textbf{Definition 2.}
The field \( b(x, t) \) propagating in the waveguide  is defined as
\begin{equation} \label{glz}
b(x, t) = \frac{1}{\sqrt{2\pi }} \int^{+\infty}_{-\infty} \mathrm{d}\omega \, b_{r}(\omega, t) e^{{\rm i}\omega \frac{x}{v_g}},
\end{equation}
where \( x \) represents the spatial coordinate along the waveguide.


Here, we consider the position $x_1$ as the origin such that we have $b(-v_gt,0)=b_{{\rm in}}(t)$. 
The waveguide field \( b(x, t) \) explicitly incorporates the spatial dependence of the traveling field through a phase factor determined by the group velocity \( v_g \).
It allows one to describe field propagation along the waveguide and to relate fields at different coupling points via finite propagation delays, which is essential for modeling spatially separated interactions.

\textbf{Theorem 2.} Given total Hamiltonian  \eqref{1}, the  input-output relation  of the giant cavity system is written as
\begin{align}\label{ar27}
b_{\rm out}(t) = b_{\rm in}(t-T) - {\rm i} \sqrt{\frac{\gamma}{2}} \big[a(t-T) + a(t)\big],
\end{align}
 where \( b_{\rm out}(t) = b(L+0^+, t) \).

\textbf{Proof.} 
Taking \eqref{ar} into \eqref{glz}, we have
\begin{eqnarray}\label{23}
b(x, t)& = & \frac{1}{\sqrt{2\pi }} \int_{-\infty}^{+\infty} \mathrm{d}\omega e^{{\rm i}\omega \frac{x}{v_g}} \bigg[ b_{r}(\omega, 0) e^{-{\rm i}\omega t} \nonumber\\
&& - {\rm i} \sum_{n=1}^2 \frac{V_q}{\sqrt{v_g}} \int_{0}^t \mathrm{d}t' a(t') e^{-{\rm i}\omega(t-t')} e^{-{\rm i}\omega \tau_n} \bigg].
\end{eqnarray}
The first term describes the free propagation of the field and can be written as
\begin{equation}\label{24}
b(x-v_g t,0)
=
\frac{1}{\sqrt{2\pi}}
\int_{-\infty}^{+\infty}\! \mathrm{d}\omega\,
b_r(\omega,0)\,
e^{\mathrm{i}\omega\frac{x-v_g t}{v_g}} .
\end{equation}
Using the identity
$
\frac{1}{2\pi}\int_{-\infty}^{+\infty} \mathrm{d}\omega\, e^{-\mathrm{i}\omega s}
= \delta(s)
$ in Eq.\eqref{23},
we have
\begin{align}
&
\int_{0}^{t}\! \mathrm{d}t'\, a(t')\,
\delta\!\left(
t-t'-\frac{x}{v_g}+\tau_n
\right)
\nonumber\\
&=
a\!\left(t-\frac{x}{v_g}+\tau_n\right)
\theta\!\left(t-\frac{x}{v_g}+\tau_n\right)
\theta\!\left(\frac{x}{v_g}-\tau_n\right).
\end{align}
Thus, one can deduce that
\begin{align}\label{25}
b(x,t)
&=
b(x-v_g t,0)
-\mathrm{i}\sqrt{\frac{\gamma}{2}}
\sum_{n=1}^2
a\!\left(t-\frac{x}{v_g}+\tau_n\right)
\nonumber\\
&\times
\theta\!\left(t-\frac{x}{v_g}+\tau_n\right)
\theta\!\left(\frac{x}{v_g}-\tau_n\right),
\end{align}
 Hence, to analyze the field at \( x = L + 0^+ \) corresponding to the waveguide output field, we have
\begin{eqnarray}
b(L+0^+, t) &=& b(L-v_g t, 0) - {\rm i} \sqrt{\frac{\gamma}{2}} \nonumber\\
&& \times\sum_{n=1}^2 a(t - \frac{L}{v_g} + \tau_n) \theta\left(\frac{L}{v_g} - \tau_n\right).\label{arL0}
\end{eqnarray}
Recalling \( \tau_1 = 0 \) and \( \tau_2 = T = \frac{L}{v_g} \), the Eq.~\eqref{arL0} can be further simplified as
\begin{equation}\label{bL0}
b(L+0^+, t) = b(L-v_g t, 0) - {\rm i} \sqrt{\frac{\gamma}{2}} \big[a(t-T) + a(t)\big].
\end{equation}
Following the above derivation, at \( x = 0^- \), the field is given by
\(
b_{\rm in}(t) = b(-v_g t, 0),
\)
so that \( b_{\rm in}(t-T)=b(L-v_g t, 0) \).
This relation explicitly connects the spatial translation of the waveguide field to a temporal delay in the effective system–field interaction.
At this point, we arrive at the input-output relation \eqref{ar27}. $\blacksquare$

\textbf{Remark 3.}
Physically, the expression \eqref{bL0} shows that the output field at position \( x=L \) and time \( t \) consists of two contributions: the freely propagating input field and the radiation emitted by the cavity at two distinct times.
The delayed term \( a(t-T) \) originates from the finite propagating time between the two coupling points, reflecting a coherent self-feedback mechanism mediated by the waveguide.
Correspondingly, the input-output relations also deviate from the standard Markovian one, where the output field $b_{\rm out}(t)$ depends on both the current and delayed  operators, as well as on a delayed input field $b_{\rm in}(t-T)$. In contrast, Markovian input-output theory yields purely instantaneous relations between input, system, and output operators. These features indicate that the giant cavity system constitutes an intrinsically non-Markovian quantum system, with dynamics and input-output structure governed by coherent time-delayed feedback rather than memoryless reservoir coupling.

%
%
\subsection{Quadrature representation}
Based on the above derivation, we can describe the cavity system coupled to a waveguide at two distant points as
\begin{align}
\dot a(t) & =\bar A a(t) +\bar A_d a(t-T)+\bar Bb_{\rm in}(t)+\bar B_db_{\rm in}(t-T),\nonumber\\
b_{\rm out}(t)& = \bar Ca(t)+\bar C_d a(t-T)+\bar D_db_{\rm in}(t-T),
\end{align}
where the coefficient matrices are written as \( \bar A = -\mathrm{i}\omega_c - \frac{\gamma}{2} \),
\( \bar A_d = -\frac{\gamma}{2} \) ,
 \(\bar  B =\bar B_d= -\mathrm{i} \sqrt{\frac{\gamma}{2}} \) ,
\( \bar C =\bar C_d= -\mathrm{i} \sqrt{\frac{\gamma}{2}} \) , and
\( \bar D_d = 1 \) .

To obtain  real coefficient equations, we convert the above equations into an quadrature representation and thus obtain
\begin{align}\label{xt}
{\rm d}x(t)
&=Ax(t)\,{\rm d}t+A_dx(t-T)\,{\rm d}t+\tilde B
\begin{bmatrix}{\rm d}w(t)\\ {\rm d}w(t-T)\end{bmatrix},\nonumber\\
{\rm d}y(t)
&=Cx(t)\,{\rm d}t+C_dx(t-T)\,{\rm d}t+D_d\,{\rm d}w(t-T),
\end{align}
which constitutes a linear Gaussian system, since the state and output dynamics are linear in the system variables and are driven by Gaussian quantum noise processes.
In the quadrature representation, the system state $x(t) = \begin{bmatrix} q(t) \ p(t) \end{bmatrix}$ constitutes the position and momentum operators of the cavity mode, where $q(t)$ and $p(t)$ are the real and imaginary parts of the mode amplitude, respectively.
And the output $\mathrm{d} y(t) = \frac{1}{\sqrt{2}}\begin{bmatrix} b_{\rm out}(t)+b^*_{\rm out}(t) \\ -i( b_{\rm out}(t)-b^*_{\rm out}(t) ) \end{bmatrix}$ and the input fields ${\rm d}{w}(t) = \frac{1}{\sqrt{2}}\begin{bmatrix} b_{\rm in}(t)+b^*_{\rm in}(t) \\ -i( b_{\rm in}(t)-b^*_{\rm in}(t) ) \end{bmatrix}$ are given in terms of their real and imaginary parts, where the signal $\rm{d} y(t)$ represents the measurement output associated with a physically observable field quadrature, rather than a direct readout of the system state.
Here, matrices $ A=\mathcal{M}(\bar A)$,
 $ A_d=\mathcal{M}(\bar A_d)$,
 $ B=\frac{1}{\sqrt{2}}\mathcal{M}(\bar B)$,
 $ B_d=\frac{1}{\sqrt{2}}\mathcal{M}(\bar B_d)$,
 $ C=\sqrt{2}\mathcal{M}(\bar C)$,
  $ C_d=\sqrt{2}\mathcal{M}(\bar C_d)$,
    $ D_d=\mathcal{M}(\bar D_d)$,
    $\tilde B= \begin{bmatrix} B & B_d \end{bmatrix}$.
The operation \( \mathcal{M}(\cdot) \) has been defined in the end of Section \ref{sec1}.

By expressing the system in the quadrature representation, we make it tractable for the design of the filter, which  estimates the system's state by processing the input and output measurements while accounting for the time delay.

\section{Optimal Filter Design}\label{section3}

In the quantum system \eqref{xt}, the system state $x(t)$ is not directly accessible, since any direct projective measurement would irreversibly disturb the quantum state. 
In practice, the available observation is obtained by continuously monitoring the output field through homodyne detection.
This measurement mechanism fundamentally distinguishes quantum filtering from its classical counterpart, as the observation process itself is constrained by quantum measurement back-action~\cite{Wiseman2010}.
More specifically, in the quadrature representation, the system state
consists of canonically conjugate operators satisfying the commutation relation $[q(t),p(t)]=\mathrm{i}$.
According to the Heisenberg uncertainty principle, such non-commuting observables cannot be simultaneously measured with arbitrary precision.
As a consequence, homodyne detection provides access only to a single output quadrature at a given time, while inevitably introducing additional uncertainty into the conjugate quadrature through measurement back-action.
This intrinsic limitation has no classical analogue and must be explicitly taken into account in the filter design~\cite{Bouten2007}.

Within this framework, the objective of quantum filtering is to construct an optimal estimator that computes the conditional expectation
\begin{align}
\hat{x}(t) = \mathbb{E}\left[ x(t) \big| \mathcal{F}_t^y  \right]
\end{align}
based on the homodyne measurement record $\{\mathrm{d} y(s):0\le s\le t\}$, while respecting both the system dynamics and the fundamental constraints imposed by quantum measurement.
Here, $\mathcal{F}_t^y$ denotes the $\sigma$-algebra generated by the homodyne observation process.
Let $e(t)=x(t)-\hat{x}(t)$ be the estimation error and $P(t)=\mathbb{E}[e(t)e^{T}(t)]$ the corresponding error covariance.
We then have the following results.

\textbf{Theorem 3.} For the giant cavity system \eqref{xt}, the optimal filter is given as 
\begin{align}\label{eq:optimal_filter}
\mathrm{d}\hat x(t)
&= A\hat x(t)\,\mathrm{d}t + A_d\hat x(t-T)\,\mathrm{d}t + \mathcal{G}(t)( D_d D_d^T)^{-1}\nonumber\\
&\qquad \times
\Bigl(
\mathrm{d}y(t)
-  C\hat x(t)\,\mathrm{d}t
-  C_d\hat x(t-T)\,\mathrm{d}t
\Bigr), 
\end{align}
where  $\mathcal{G}(t)=P(t) C^T+P_1(t)C_d^T$ with the covariance matrix $P(t)$ defined as
$P(t)=\mathbb{E}[(x(t)-\hat x(t))(x(t)-\hat x(t))^T|\mathcal{F}_t^y]$
 and $P_{1}(t) =
\begin{cases}
\mathbb{E}[e(t)e^T(t-T)|\mathcal{F}_t^y], & t\geq T\\
0, & t<T.
\end{cases}$ is the delayed
error covariance matrix.

\textbf{Proof.}
We first define the predicted output
$
\mathrm{d}\hat y(t)= C\hat x(t)\mathrm{d}t+ C_d\hat x(t-T)\mathrm{d}t,
$
and introduce the innovation process
\begin{equation}\label{eq:innovation}
\mathrm{d}\nu (t)
= \mathrm{d} y(t)-\mathrm{d}\hat y(t).
\end{equation}
Since $ D_d D_d^T$ is positive definite, $\nu(t)$ is an $\mathcal Y_t$-martingale
with incremental covariance
$\mathbb E[\mathrm{d}\nu(t)\mathrm{d}\nu^T(t)]
= D_d D_d^T\,\mathrm{d}t.$
For linear Gaussian system \eqref{xt}, the minimum mean square error filter admits the structure
\begin{equation}\label{eq:filter_general}
\mathrm{d}\hat x(t)
= A\hat x(t)\,\mathrm{d}t + A_d\hat x(t-T)\,\mathrm{d}t
+ K(t)\,\mathrm{d}\nu(t),
\end{equation}
where $K(t)$ denotes the filter gain.
Substituting \eqref{eq:innovation} into \eqref{eq:filter_general} yields
\begin{align}\label{eq:filter_struct}
\mathrm{d}\hat x(t)
&= A\hat x(t)\,\mathrm{d}t + A_d\hat x(t-T)\,\mathrm{d}t \\
&\quad + K(t)\Bigl(
\mathrm{d}y(t)
-  C\hat x(t)\,\mathrm{d}t
-  C_d\hat x(t-T)\,\mathrm{d}t
\Bigr).\nonumber
\end{align}
The optimality condition follows from the orthogonality principle 
$\mathbb E[e(t)\,\mathrm{d}\nu^T(t)]=0$.
Since the innovation process $\mathrm{d}\nu$ is an $\mathcal{F}_t^y$-Brownian motion, the observation space admits the orthogonal decomposition
\begin{align}
L^2(\mathcal{F}_t^y)
=
L^2(\mathcal{F}_{t-}^y)
\oplus \mathrm{span}\{\mathrm{d}\nu(t)\}.
\end{align}
By the Hilbert space projection theorem \cite{Maybeck1979}, 
the estimator admits $\hat x(t)=\hat x(t^-)+K(t)\mathrm d\nu(t)$.
The orthogonality principle then implies
$\mathbb E[(x(t)-\hat x(t))\mathrm d\nu^T(t)]=0$,
which yields
$\mathbb E[x(t)\mathrm d\nu^T(t)]=K(t)\mathbb E[\mathrm d\nu(t)\mathrm d\nu^T(t)]$.
Hence,
$K(t)=\mathbb E[x(t)\mathrm d\nu^T(t)]\big(\mathbb E[\mathrm d\nu(t)\mathrm d\nu^T(t)]\big)^{-1}$.
Since $\mathbb E\!\left[x(t)\,\mathrm{d}\nu^T(t)\right]=[P(t) C^T+P_1(t)C_d^T]\mathrm{d}t$,
 the optimal filter gain is given as
\begin{equation}\label{eq:gain}
K(t)=[P(t) C^T+P_1(t)C_d^T]( D_d D_d^T)^{-1}.
\end{equation}
Substituting \eqref{eq:gain} into \eqref{eq:filter_struct}, the optimal
filter is given by Eq. \eqref{eq:optimal_filter}. $\blacksquare$

\textbf{Remark 4.}
In contrast to the approach in Ref.~\cite{Basin}, where the optimal filter gain depends solely on the covariance matrix $P(t)$, the optimal filter gain in \eqref{eq:optimal_filter} involves both $P(t)$ and $P_1(t)$. This structural difference is a direct consequence of the output delay inherent in system \eqref{xt}, which introduces additional correlation terms between the current state and delayed states that must be explicitly accounted for in the filter design.
The term $ D_d D_d^T=I$ can be obtained from Subsection 2.4.
The time evolution of estimation error $P(t)$ can be deduced as follows.

\textbf{Theorem 4.}
Given the linear quantum stochastic delay system \eqref{xt}
and the filter \eqref{eq:optimal_filter},
the conditional error covariance $P(t)$ satisfies
\begin{align}\label{eq:Pdot_correct_align}
\dot P(t)
&=(A-\mathcal G(t)C)\,P(t)+P(t)\,(A-\mathcal G(t)C)^T \nonumber\\
&\quad+(A_d-\mathcal G(t)C_d)\,P_1^T(t)+P_1(t)\,(A_d-\mathcal G(t)C_d)^T \nonumber\\
&\quad+BB^T
+\bigl(B_d-\mathcal G(t)D_d\bigr)\bigl(B_d-\mathcal G(t)D_d\bigr)^T,
\end{align}
with $P_1(t)=0$ for $t<T$.

\textbf{Proof.}
The error dynamics is
\begin{align}\label{eq:error_dyn_align}
{\rm d}e(t)
&={\rm d}x(t)-{\rm d}\hat x(t)\nonumber\\
&=\Bigl((A-\mathcal G(t)C)e(t)+(A_d-\mathcal G(t)C_d)e(t-T)\Bigr){\rm d}t
\nonumber\\&
+B\,{\rm d}w(t)+\bigl(B_d-\mathcal G(t)D_d\bigr){\rm d}w(t-T).
\end{align}
Applying the It\^o product rule to $e(t)e^T(t)$ gives
\begin{align}
{\rm d}\bigl(e(t)e^T(t)\bigr)=({\rm d}e(t))e^T(t)+e(t)({\rm d}e(t))^T+({\rm d}e(t))({\rm d}e(t))^T.
\end{align}
Taking $\mathbb E[\cdot\mid\mathcal F_t^y]$ on both sides yields
\begin{align}\label{eq:dP_split_align}
{\rm d}P(t)= {\rm d}R_t+{\rm d}S_t+{\rm d}Q_t,
\end{align}
where
${\rm d}R_t=\mathbb E\![({\rm d}e(t))e^T(t)$  
$|\mathcal F_t^y]$,
${\rm d}S_t=\mathbb E\![e(t)({\rm d}e(t))^T$  $\mid\mathcal F_t^y]$,
${\rm d}Q_t=\mathbb E\!\left[({\rm d}e(t))({\rm d}e(t))^T\mid\mathcal F_t^y\right]$.
Using \eqref{eq:error_dyn_align} and $\mathbb E[{\rm d}W(\cdot)\mid\mathcal F_t^y]=0$,
\begin{align}
{\rm d}R_t
&=\Bigl((A-\mathcal G(t)C)\,P(t)+(A_d-\mathcal G(t)C_d)\,P_1^T(t)\Bigr){\rm d}t,\\
{\rm d}S_t
&=\Bigl(P(t)\,(A-\mathcal G(t)C)^T+P_1(t)\,(A_d-\mathcal G(t)C_d)^T\Bigr){\rm d}t.
\end{align}
Since Wiener increments satisfy ${\rm d}W(t){\rm d}W^T(t)=I_m{\rm d}t$ and
${\rm d}W(t)$ is independent of ${\rm d}W(t-T)$ for $T>0$,
the cross-variation between the two noise channels vanishes. Therefore,
\begin{align}
{\rm d}Q_t
&=\Bigl(BB^T
+\bigl(B_d-\mathcal G(t)D_d\bigr)\bigl(B_d-\mathcal G(t)D_d\bigr)^T\Bigr){\rm d}t.
\end{align}
Combining ${\rm d}R_t$, ${\rm d}S_t$, and ${\rm d}Q_t$ in \eqref{eq:dP_split_align} 
gives \eqref{eq:Pdot_correct_align}. \hfill$\blacksquare$

\textbf{Remark 5.}
From Theorem~4, it can be observed that when the current  time \(t\) is less than the time delay \(T\), \(\mathrm{d}P(t)\) can be expressed directly. However, when \(t\) exceeds \(T\), the expression of \(\mathrm{d}P(t)\) depends both on the information of \(P_1(t)\) and $P(t)$. Consequently, determining how to compute \(P_1(t)\) for \(t \geq T\), and then successively derive \(P_j(t)\) for \(t \geq jT\), becomes a crucial step in designing the filter.

Next, we present the derivation for calculating the delay state covariance matrix.

\textbf{Theorem 5.}
Given the linear quantum stochastic delay system \eqref{xt}, the filter \eqref{eq:optimal_filter}, $P(0)=\mathbb E[e(0)e^T(0)]$ and $P_j(t)=0$ for $t<jT$,
define for $j\in\mathbb N^+$
\begin{align}
P_j(t)=
\begin{cases}
\mathbb E\!\left[e(t)e^T(t-jT)\mid\mathcal F_t^y\right], & t\ge jT,\\
0, & t<jT,
\end{cases}
\end{align}
where $e(t)=x(t)-\hat x(t)$.
Then, for $t\in[jT,+\infty)$, $P_j(t)$ satisfies
\begin{align}\label{eq:Pj_correct}
\dot P_j(t)
&=(A-\mathcal G(t)C)\,P_j(t)
+(A_d-\mathcal G(t)C_d)\,P_{j-1}(t-T)\nonumber\\
&\quad+P_j(t)\,(A-\mathcal G(t-jT)C)^T
\nonumber\\
&\quad+P_{j+1}(t)\,(A_d-\mathcal G(t-jT)C_d)^T\nonumber\\
&\quad+\bigl(B_d-\mathcal G(t)D_d\bigr)B^T\,\mathbb I_{j=1},
\end{align}
with the convention $P_0(t)=P(t)$ and $P_{j-1}(t-T)=0$ if $t-T<(j-1)T$.

\textbf{Proof.}
For $t\ge jT$, applying the It\^o product rule to $e(t)e^T(t-jT)$ yields
\begin{align}\label{eq:ito_Pj}
{\rm d}\!\bigl(e(t)e^T(t-jT)\bigr)
&=({\rm d}e(t))e^T(t-jT)+e(t)({\rm d}e(t-jT))^T
\nonumber\\
&+({\rm d}e(t))({\rm d}e(t-jT))^T .
\end{align}
Taking $\mathbb E[\cdot\mid\mathcal F_t^y]$ on both sides gives
\begin{align}\label{eq:dPj_split}
{\rm d}P_j(t)={\rm d}R_j(t)+{\rm d}S_j(t)+{\rm d}Q_j(t),
\end{align}
where
${\rm d}R_j(t)=\mathbb E\!\left[({\rm d}e(t))e^T(t-jT)\mid\mathcal F_t^y\right]$,
${\rm d}S_j(t)=\mathbb E\!\left[e(t)({\rm d}e(t-jT))^T\mid\mathcal F_t^y\right]$,
${\rm d}Q_j(t)=\mathbb E\![({\rm d}e(t))($ ${\rm d} e(t-jT))^T\mid\mathcal F_t^y]$.

Using the error dynamics \eqref{eq:error_dyn_align} in Theorem~3 and $\mathbb E[{\rm d}W(\cdot)\mid\mathcal F_t^y]=0$,
\begin{align}\label{eq:dRj}
{\rm d}R_j(t)
&=\Bigl((A-\mathcal G(t)C)\,P_j(t)
+(A_d-\mathcal G(t)C_d)\,\nonumber\\
&\quad \times \mathbb E[e(t-T)e^T(t-jT)\mid\mathcal F_t^y]\Bigr){\rm d}t\nonumber\\
&=\Bigl((A-\mathcal G(t)C)\,P_j(t)
+(A_d-\mathcal G(t)C_d)\,
\nonumber\\
&\quad \times P_{j-1}(t-T)\Bigr){\rm d}t,
\end{align}
where we used $e(t-jT)=e((t-T)-(j-1)T)$ and thus
$\mathbb E[e(t-T)e^T(t-jT)\mid\mathcal F_t^y]=P_{j-1}(t-T)$.
Applying \eqref{eq:error_dyn_align} at time $t-jT$ yields
\begin{align}
{\rm d}e(t-jT)
&=\Bigl((A-\mathcal G(t-jT)C)e(t-jT)\nonumber\\
&\quad+(A_d-\mathcal G(t-jT)C_d)e(t-(j+1)T)\Bigr){\rm d}t\nonumber\\
&\quad+B\,{\rm d}w(t-jT)+\bigl(B_d-\mathcal G(t-jT)D_d\bigr)
\nonumber\\
&\quad \times {\rm d}w(t-(j+1)T).
\end{align}
Taking conditional expectation and using $\mathbb E[{\rm d}w(\cdot)\mid\mathcal F_t^y]=0$ again,
\begin{align}\label{eq:dSj}
{\rm d}S_j(t)
&=\Bigl(P_j(t)\,(A-\mathcal G(t-jT)C)^T
\nonumber\\
&\quad+\mathbb E[e(t)e^T(t-(j+1)T)\mid\mathcal F_t^y]\,\nonumber\\
&\quad\times (A_d-\mathcal G(t-jT)C_d)^T\Bigr){\rm d}t\nonumber\\
&=\Bigl(P_j(t)\,(A-\mathcal G(t-jT)C)^T
\nonumber\\
&\quad+P_{j+1}(t)\,(A_d-\mathcal G(t-jT)C_d)^T\Bigr){\rm d}t.
\end{align}
By the It\^o rule, ${\rm d}Q_j(t)$ is determined by the quadratic covariation
between the diffusion terms in ${\rm d}e(t)$ and ${\rm d}e(t-jT)$.
Since Wiener increments at distinct times are independent,
${\rm d}w(t)$ is independent of ${\rm d}w(t-jT)$ and ${\rm d}w(t-(j+1)T)$, and
${\rm d}w(t-T)$ coincides with ${\rm d}w(t-jT)$ if and only if $j=1$.
Therefore,
\begin{align}\label{eq:dQj}
{\rm d}Q_j(t)=\bigl(B_d-\mathcal G(t)D_d\bigr)B^T\,\mathbb I_{j=1}\,{\rm d}t.
\end{align}
Combining \eqref{eq:dRj}--\eqref{eq:dQj} in \eqref{eq:dPj_split} 
gives \eqref{eq:Pj_correct}. \hfill$\blacksquare$

Based on the above, we present the optimal filter design algorithm for system \eqref{xt}.

%

 \noindent\textbf{Algorithm 1: Optimal Filter Design for the cavity coupled to  the waveguide at two distant points }

\noindent\textbf{Input:} Initial state $ x(0)$, time delay $T$.

\noindent\textbf{Output:} Stochastic state estimate $\hat x(t)$, which represents the conditional expectation of the system state.

\textbf{Step 1.} Initialize the filter state $\hat x(0)=\hat x_0$ for $t\in[-T,0]$. Set $m=\left\lfloor \frac{t}{T} \right\rfloor $;

\textbf{Step 2.} A terminal condition $m=0$ is given. If this condition is satisfied, go to \textbf{Step 4}; if not, go to \textbf{Step 3};

\textbf{Step 3.} Compute the delayed state covariance matrices $\mathrm{d}P_m(t)$ using~\eqref{eq:Pj_correct}. Set $m=m-1$. Return to \textbf{Step 2};

\textbf{Step 4.} Derive the differential equation of the covariance matrix $\mathrm{d} P(t)$ with \eqref{eq:Pdot_correct_align}. Compute the optimal filter \eqref{eq:optimal_filter}.


Algorithm~1 presents a recursive procedure for the optimal filter design of a cavity coupled to a waveguide at two distant points. 
The algorithm proceeds by partitioning the time axis into delay intervals of length~$T$ and recursively propagating the delayed cross-covariance matrices $P_j(t)$. 
This interval-wise backward recursive structure explicitly captures the effect of mixed delays on the estimation error dynamics.

%
%
%
%

\section{Examples}\label{section4}
In this section, we will demonstrate the effectiveness of our proposed filter design method through simulations. Specifically, we will illustrate the estimation performance of the filter using coherent and cat states. Wigner function is a quasi-probability distribution used in quantum mechanics to represent quantum states in phase space. It provides a complete description of a quantum system and is useful for visualizing quantum states and their properties in the position \( (q) \) and momentum \( (p) \) coordinates.

 \subsection{Estimation of coherent states}
 The Wigner function for a coherent state  is characterized by  Gaussian distribution in phase space.
The coherent state's Wigner function centered at phase-space coordinates $(\langle q \rangle, \langle p \rangle)$ is given by
\begin{equation}
W(q,p) = \frac{2}{\pi} \exp\left(-2\left[(q - \langle q \rangle)^2 + (p - \langle p \rangle)^2\right]\right).
\end{equation}
This expression describes a symmetric Gaussian distribution in the phase-space variables $q$ (position) and $p$ (momentum), where $\langle q \rangle$ and $\langle p \rangle$ represent the expectation values of position and momentum for the coherent state.

We validate our proposed filter design approach through a number of simulations with the following parameters: frequency $\omega_c=10^9\rm{Hz}$, decay  rate 
$\gamma=8\times10^8\rm{Hz}$, group velocity $v_g=10^3\rm{m/s}$. The initial states of the original system and filter are given as $[-4\rm{m}, 4\rm{kg\cdot m/s}]^T$ and $[4\rm{m},-4\rm{kg\cdot m/s}]^T$, respectively. We set the time delay  as $T=1.5\times10^{-8}\rm{s}$, in other words, $L=1.5\times10^{-5}\rm{m}$.

\begin{figure}[] 
  \centering
  \includegraphics[width=8cm]{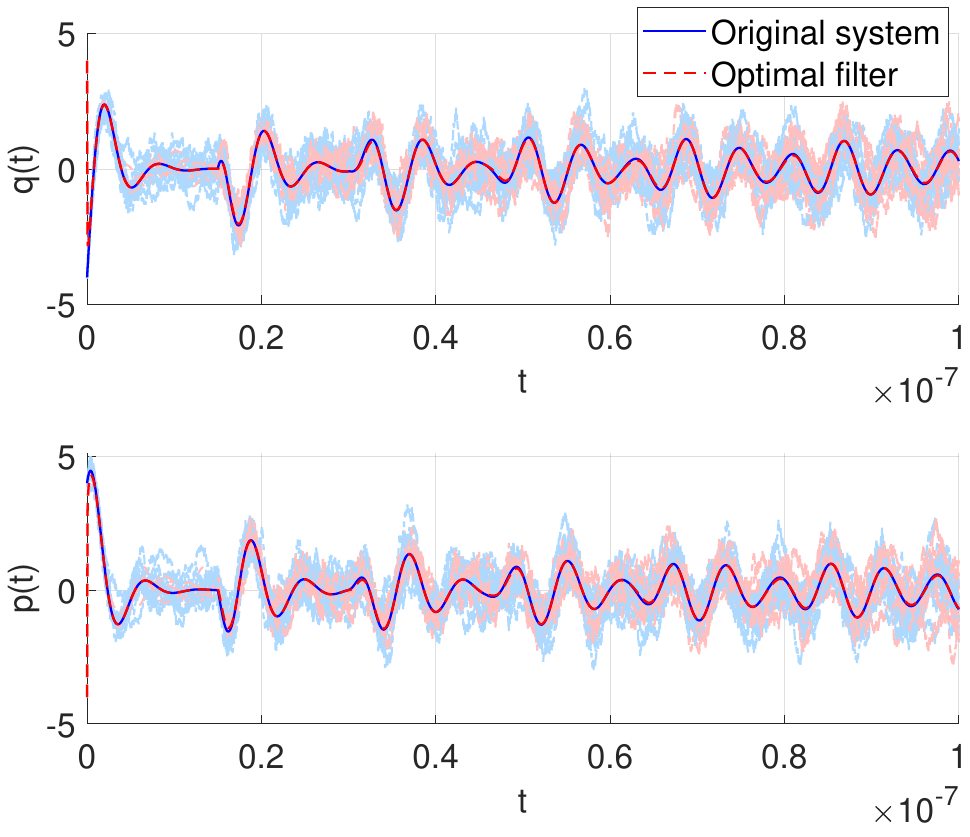} 
  \caption{Dynamical evolution of position and momentum operators of the system and filter}
  \label{fig_state}
\end{figure}

\begin{figure*}[t] 
  \centering
  \includegraphics[width=18cm]{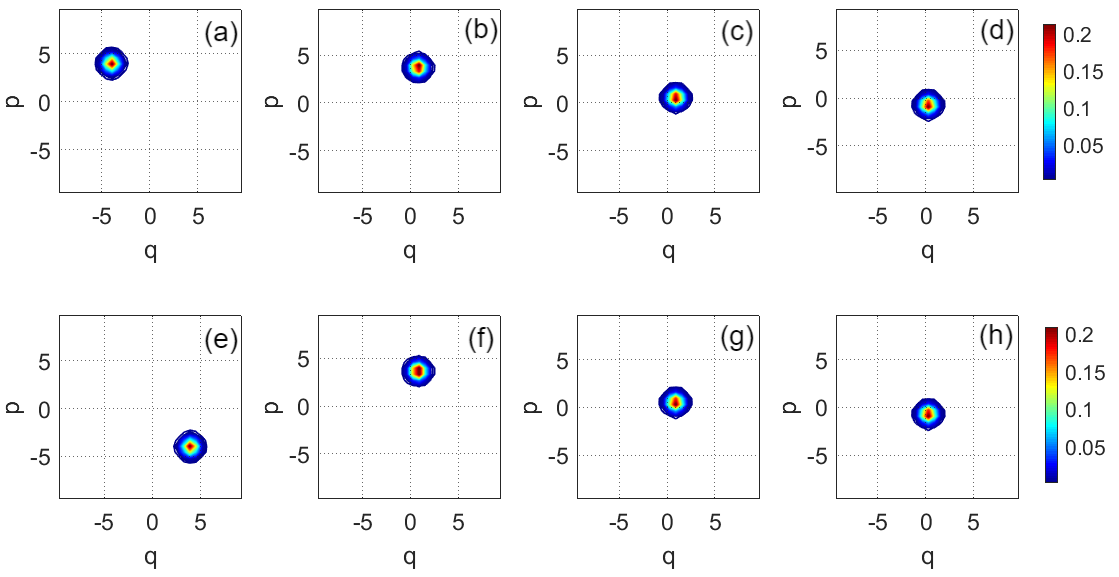}
  \caption{Schematic of Wigner functions for coherent states in different moments.
Figures (a)-(d) depict the evolution of the coherent state in the original system: (a) at $t = 0$, (b) at $t = 0.01T_f$, (c) at $t = 0.5T_f$, and  (d) at $t = T_f$.
Figures (e)-(h) illustrate the corresponding evolution of the coherent state in the filter system: (e) at $t = 0$, (f) at $t = 0.01T_f$, (g) at $t = 0.5T_f$ and (h) at $t = T_f$.}
  \label{fig_wide}
\end{figure*}

In Fig. \ref{fig_state}, we illustrate the tracking performance of the filter state relative to the original system state over the time interval $[0, 1\times10^{-7}\rm{s}]$.
The light blue dashed lines represent multiple random state trajectories of the original system under the influence of noise, thereby reflecting the fluctuation range of the true system states under noise perturbations. The dark blue curve denotes the mean trajectory of the original system, illustrating the overall evolutionary trend of the system states. The light red dashed lines indicate multiple estimated trajectories obtained by the filter under the same conditions, while the red dashed line represents the mean estimated trajectory of the filter states.
It should be emphasized that the envelope of the curves exhibits oscillations around $1.5\times 10^{-8}\rm{s}$  indicating non-Markovian dynamics, which is different from monotonical decays in the Markovian case.
It can be observed that, despite a significant initial discrepancy between the filter's initial state and that of the original system, the filter state rapidly converges to and tracks the system state. Fig. \ref{fig_wide} displays the coherent states of the original system and the filter at different time instants. Specifically, subfigures (a) and (e) present the coherent states of the original system and the filter at the initial time, showing a clear difference between them. However, as time progresses at instants $0.01T_f$ and $0.5T_f$, the filter state gradually approaches that of the original system. By time $T_f$, the coherent state of the filter closely matches that of the source system.

\begin{figure}[] 
  \centering
  \includegraphics[width=9cm]{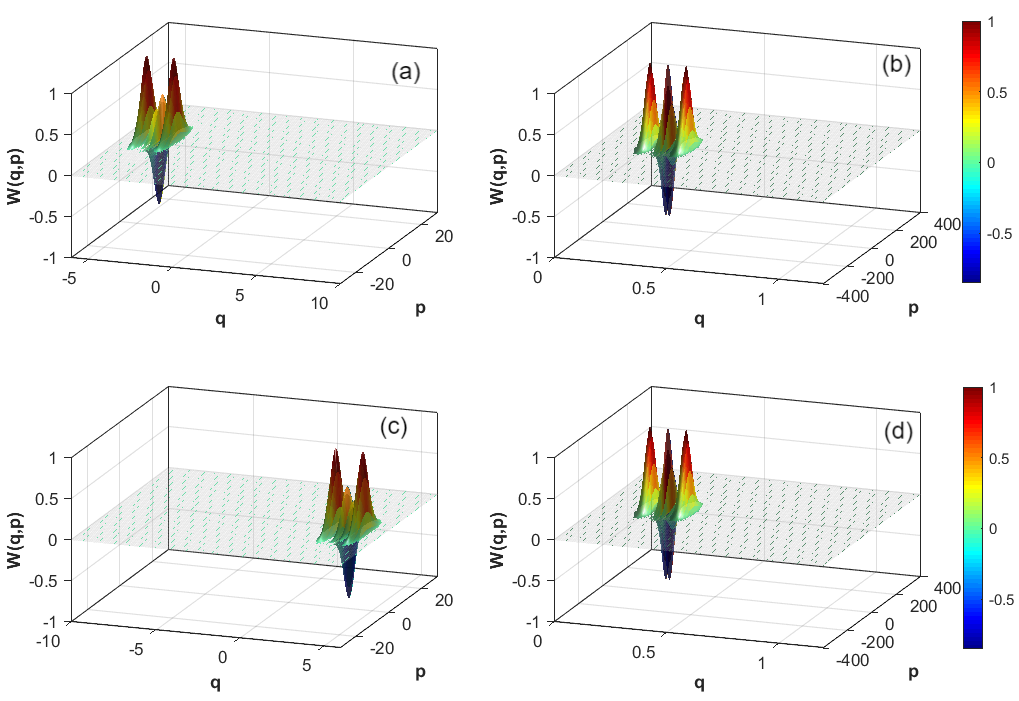} 
  \caption{Wigner function of the cat state at different time points.
Figures (a)-(b) show the evolution of the cat state in the original system: (a) at $t = 0$ and (b) at $t = T_f$.
Figures (c)-(d) display the corresponding evolution in the filter system: (c) at $t = 0$ and (d) at $t = T_f$.
For Fig. (a), $q_0=-4$, $p_0=4$, $\beta = 0.8$ and $\sigma = 0.2$; for Fig. (b), $q_0=0.29$, $p_0=-0.71$, $\beta = 0.08$ and $\sigma = 0.02$;
for Fig. (b), $q_0=4$, $p_0=-4$, $\beta = 0.8$ and $\sigma = 0.2$;
for Fig. (b), $q_0=0.29$, $p_0=-0.68$, $\beta = 0.08$ and $\sigma = 0.02$.}
  \label{fig_cat}
\end{figure}

\subsection{Estimation of cat states}

In addition, the Wigner function provides a complete phase-space representation of quantum states and is particularly effective in characterizing non-classical features of Schr\"{o}dinger cat states. Such states are coherent superpositions of macroscopically distinct coherent states and exhibit pronounced quantum interference effects in phase space.

Specifically, we consider an even Schr\"{o}dinger cat state formed by two coherent states $|\alpha\rangle$ and $|-\alpha\rangle$, defined as
\begin{equation}
|\mathrm{cat}\rangle
=
\frac{1}{\sqrt{2\!\left(1+e^{-2|\alpha|^2}\right)}}
\left(|\alpha\rangle+|-\alpha\rangle\right),
\end{equation}
where $\alpha\in\mathbb{C}$ denotes the coherent-state amplitude. The corresponding density operator is given by
$\rho = |\mathrm{cat}\rangle\langle \mathrm{cat}|$.
To evaluate the Wigner function, we first express the coherent states in the Fock basis as
\begin{equation}
|\alpha\rangle
=
e^{-|\alpha|^2/2}
\sum_{n=0}^{\infty}
\frac{\alpha^n}{\sqrt{n!}}\,|n\rangle.
\end{equation}
Motivated by the closed-form expression of the Wigner function for a superposition
of two displaced coherent states, the Wigner function is decomposed into three parts:
two Gaussian contributions associated with the individual coherent states and an
interference term capturing quantum coherence \cite{Schleich2015}.
Specifically, the two coherent components are given by
\begin{align}
W_{\mathrm{coh}}^{\pm}(q,p)
=
\exp\!\left(
-\frac{(q-q_0\mp\beta)^2}{\sigma^2}
-\sigma^2 (p-p_0)^2
\right),
\end{align}
where $(q_0,p_0)$ denotes the phase-space center of the cat state, $\beta$ represents
the separation between the two coherent components, and $\sigma>0$ characterizes the
effective width of each Gaussian packet.
The specific values of $\sigma$ and $\beta$ are provided in the caption of Fig. \ref{fig_cat}.

The quantum interference between the two coherent components is described by
\begin{align}
W_{\mathrm{int}}(q,p)
&=
\exp\!\left(
-\frac{(q-q_0)^2}{\sigma^2}
-\sigma^2 (p-p_0)^2
\right)\nonumber\\
&\times
\cos\!\left(
\frac{2\beta}{\sigma^2}(q-q_0)
-
2\beta\sigma^2(p-p_0)
\right),
\end{align}
which generates the characteristic oscillatory fringes in phase space.

Combining these terms, the Wigner function used in the numerical simulations is
approximated by
\begin{equation}
\bar W(q,p)
=
\frac{1}{2}
\left(
W_{\mathrm{coh}}^{+}(q,p)
+
W_{\mathrm{coh}}^{-}(q,p)
+
W_{\mathrm{int}}(q,p)
\right),
\end{equation}
followed by a normalization
\begin{equation}
 W(q,p)=\frac{\bar W(q,p)}{\max_{q,p}| \bar W(q,p)|},
\end{equation}
which is introduced solely for visualization purposes and does not affect the
underlying phase-space structure.

The resulting Wigner function for cat states typically shows an interference pattern between the two coherent state components, with characteristic negative regions that reveal the quantum nature of the superposition, as shown in Fig. \ref{fig_cat}. This negativity distinguishes cat states from classical mixtures and demonstrates their non-classical behavior.
In Fig. \ref{fig_cat}, subfigures (a) and (c) present the initial cat states of the original system and the filter, respectively. As indicated by their Wigner functions, there is a significant difference between the two initial cat states. However, by the final time \( T_f \), the cat states shown in subfigures (b) and (d) demonstrate that the filter provides a highly accurate estimate of the original system's state.

\section{Conclusion}\label{section5}
This paper systematically studies the optimal filtering for non-Markovian cavity in waveguide QED systems.
Unlike the commonly used Markovian methods in existing works, we  formulate the Langevin equation of such  quantum systems in the Heisenberg picture as a linear continuous system with mixed delays. This approach fully preserves the intrinsic dynamic characteristics arising from non-local coupling and multiple propagation delays at the modeling level, thereby avoiding excessive simplification of the original system's physical structure.
Building on this,  this paper establishes a systematic theoretical framework for optimal filtering, centered on characterizing the dynamic evolution of the error covariance matrix. Thus,  an optimal filter design algorithm for such quantum systems is presented through an
interval-wise recursive procedure.
Numerical simulation results verify that the proposed method effectively enhances state estimation performance while preserving the physical characteristics of the system. 
Future work will extend this filtering framework from the single-cavity two-point coupling case to more complex quantum network systems with multi-point and multi-cavity couplings.

\end{document}